\documentstyle[aps,twocolumn,epsfig,rotating]{revtex}

\begin{document} 

\draft
\twocolumn[\hsize\textwidth\columnwidth\hsize\csname    
@twocolumnfalse\endcsname                               

\begin{title}
		{Muon Spin Relaxation Study of Reentrant Spin Glasses: 
			  Amorphous-Fe$_{1-x}$Mn$_x$}

\end{title}

\author{
		  M.J.P. Gingras$^{1}$, M. Larkin$^2$, I. Mirebeau$^3$, W.D. Wu$^2$,
		  K.  Kojima$^2$, G.M.   Luke$^2$,
		   B.  Nachumi$^2$,
		  and Y.J.  Uemura$^2$        }

\address{$^1$Department of Physics, University of
Waterloo, Waterloo, ON, N2L 3G1, Canada}
\address{$^2$ Department of Physics, Columbia University,
New York, NY 10027, USA} 
\address{$^3$ Laboratoire L\'eon Brillouin,
CEN-Saclay, 91191 Gif-sur-Yvette, cedex, France} 

\date{\today}

\maketitle

\begin{abstract}
\noindent 
We have performed an 
extensive muon spin relaxation study of the well known
(Fe$_{1-x}$Mn$_{x}$)$_{75}$P$_{16}$B$_6$Al$_3$ reentrant
spin glass.
We find a strong increase of the dynamic as well
as static muon depolarization rate at the paramagnetic to ferromagnetic
transition temperature, $T_c$, and at $T_f$ ($T_f<T_c$) that
corresponds to the onset of strong bulk magnetic irreversabilities. We
find no {\it critical dynamic} signature 
of a freezing of the transverse $XY$ spin
components at an intermediate temperature $T_{xy}$ ($T_f < T_{xy}  <
T_c$) where extra static order not contributing to the magnetization
starts developping.
\end{abstract}


\vskip2pc]                                              

\narrowtext

\noindent Frozen random disorder added to competing interactions leads
to random frustration in condensed matter systems~\cite{Toulouse}, a feature
that is detrimental to the stability of the long-range ordered state
otherwise present in the disorder-free pure material.  Random
frustration is ubiquitous in condensed matter physics~\cite{Toulouse};
it arises in random magnets~\cite{BYFH}, disordered Josephson junction
arrays in a magnetic field~\cite{JJ}, mixed molecular
systems~\cite{OG}, and
in partially UV polymerized lipid membranes~\cite{membranes}.
One expects  these systems to share similar thermodynamic behavior and,
in particular, to display a long-range ordered phase with
low-temperature glasslike properties~\cite{Toulouse,BYFH}.
Hence, an improved understanding of the simpler randomly frustrated ferromagnets
would contribute to our overall understanding of several other types of
weakly frustrated systems.   

The mean-field theory of weakly randomly frustrated Heisenberg
ferromagnets predicts a sequence of {\it three}
transitions~\cite{BYFH}.  Upon cooling, a transition from a
paramagnetic phase to a collinear ferromagnetic state
first occurs at $T_c$, below which the system develops a nonzero
bulk magnetization,
$M$ (e.g. along the $\hat z$ direction).
At a lower temperature, $T_{xy}$,  the transverse $XY$ spin
components perpendicular to $M$ freeze in random directions, with no
decrease of $M$ at $T<T_{xy}$.    The
temperature vs disorder, $x$,  boundary, $T_{xy}(x)$,  is usually referred
to as the Gabay-Toulouse line~\cite{BYFH}.  Finally, strong
longitudinal irreversibilities develop at a third temperature, $T_f$
($T_f<T_{xy}$), with again no decrease of 
$M$ at $T\le T_f$~\cite{Ryan,Mirebeau,Gingras-1,Gingras-2}.
The boundary $T_f(x)$
in the mean-field theory of Heisenberg spin glasses 
is a remnant of the longitudinal freezing seen
below the Almeida-Thouless line in weakly frustrated  Ising spin
glasses~\cite{BYFH}.  In mean-field theory, $T_{xy}$ is a true
thermodynamic (transverse spin-glass) phase transition, while $T_f$ is
a sharp crossover temperature rather than a transition per se.  Systems
exhibiting such {\it mixed} ferromagnetic and spin glass behavior are
commonly referred to as ``reentrant spin
glasses''~\cite{Ryan,Mirebeau,Gingras-1,Gingras-2}.  

An extensive experimental effort using macroscopic (AC and DC magnetization) and
microscopic (neutrons, M\"ossbauer spectroscopy and EPR) probes have
been used in the past twenty years 
to search for the above three transitions in real materials,
and to determine their nature~\cite{Ryan,Mirebeau,Gingras-1}.  The
ferromagnetic and freezing transitions at $T_{\rm c}$ and $T_f$ are
most easily observed through anomalies of the magnetic
susceptibility~\cite{BYFH,Ryan,Mirebeau,Gingras-1}.  Signatures of the
transverse freezing transition at $T_{xy}$ have apparently been seen in
Au$_{81}$Fe$_{19}$, via  magnetic Bragg peak intensity~\cite{Murani}
and EPR~\cite{Coles} measurements, in amorphous
Fe$_{100-x}$Zr$_x$~\cite{Ryan} and
in (Fe$_{0.65}$Ni$_{0.35}$)$_{1-x}$Mn$_x$~\cite{Huck} using M\"ossbauer
spectroscopy, and in amorphous
(Fe$_{1-x}$Mn$_{x}$)$_{75}$P$_{16}$B$_6$Al$_3$ \hspace{2truept}
(a$-$Fe$_{1-x}$Mn$_x$) using neutron depolarization, small angle
neutron scattering and inelastic neutron scattering~\cite{Mirebeau}.
However, there is at present very little evidence for a  clear
signature of these three transitions as demonstrated via 
a {\it single} experimental
technique on a  {\it given} material.   An exception may be the
experiments mentioned above on Au$_{81}$Fe$_{19}$~\cite{Murani,Coles},
where three ``features'' that may be associated with $T_{\rm c}$,
$T_{xy}$ and $T_f$ have been seen.  This raises the long standing and
still main question at stake in the problem of reentrant spin glasses:

\vspace{1mm}
\begin{center}
{\sl In real materials, as opposed to mean-field theory, are  $T_{xy}$
and $T_f$ really two distinct transitions or thermodynamic features,
or, are they
the same one observed at
different time and length scales?}
\end{center}
\vspace{1mm}

To  help resolve this issue, we have performed an extensive $\mu$SR study of 
the well known a$-$Fe$_{1-x}$Mn$_x$ reentrant ferromagnetic spin glass 
material~\cite{Mirebeau}. $\mu$SR is an ideally suited technique to study the
above question
as it allows to extract {\it simultaneously} 
the amount of static spin order 
and to monitor the level of the spin dynamics.
Specifically, we have searched 
for three well defined signatures of transitions at $T_{\rm c}$, $T_{xy}$ and 
$T_f$ in the muon relaxation spectrum. 
Barsov et al.~\cite{barsov} performed earlier $\mu$SR study in
Fe$_{x}$Ni$_{80-x}$Cr$_{20}$ and (Fe$_{x}$Mn$_{1-x}$)Pt$_{3}$ alloys.
They found no signature of anomaly in dynamic muon spin relaxation rate
$\lambda$
corresponding to $T_{xy}$ in the former system, while some broad feature
which might be related to $T_{xy}$, has been seen in the latter.  Their results were,
however, not quite conclusive, due to very small variable range of
$\lambda \leq 1.5 \mu s^{-1}$ in their data and to the lack of
results for the contribution from the static depolarization rate.

In order to build a clear comparison between real systems and theory, we
have focused our experimental efforts on an {\it archetype} 
reentrant 
ferromagnetic spin glass material, such as  a$-$Fe$_{1-x}$Mn$_x$,  which 
has been investigated by several other techniques~\cite{Mirebeau}. 
The disorder-temperature 
phase diagram of a$-$Fe$_{1-x}$Mn$_x$ is shown in the inset panel of Fig.1).
  We have investigated reentrant spin glass samples with 
concentration $x=0.26$ and $0.30$, and a strongly frustrated sample, $x=0.41$, 
which shows full blown spin-glass behavior~\cite{BYFH,Mirebeau}. 
\begin{figure}
\begin{center}
  {
  \begin{turn}{0}%
    {\epsfig{file=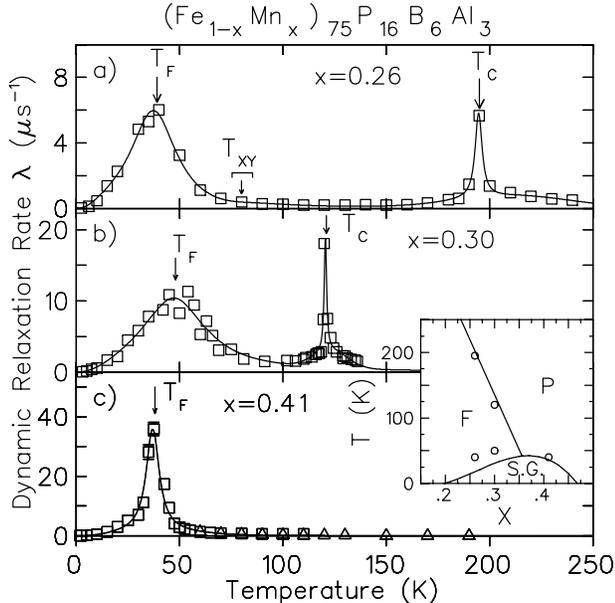,height=8cm,width=8cm} }
   \end{turn}
   }
\caption{Temperature dependence of the dynamical
relaxation rate, $\lambda(T)$, for the reentrant samples,
$x=0.25$ (Fig. 1a) and $x=0.30$ (Fig. 1b), and for the
spin glass sample, $x=0.41$ (Fig. 1c).
The inset panel shows the $T-x$ phase diagram for
a-Fe$_{1-x}$Mn$_x$ from DC magnetization measurements (solid lines).
The open circles are the location of the
peaks in the dynamical relaxation rate, $\lambda$, obtained in this
$\mu$SR experiment.}
\end{center}
\end{figure}
The experiments were performed on beamlines M13 and M20 at TRIUMF.  
In essence, in zero-field (ZF) and
longitudinal field (LF) $\mu$SR measurements, the positive muon ($\mu^+$)
beam, spin polarized parallel to the incoming beam direction, stops in
the sample, and time histograms of muon decay positrons are recorded by
forward, $F$, and and backward, $B$, counters as a function of the time of
residence, $t$, of each $\mu^+$ in the specimen.  Since a positron is
emitted preferentially parallel to the instantaneous $\mu^+$ spin
direction, the asymmetry $G(t)= (F-B)/(F+B)$ is a  statistical measure
(in the quantum mechanical sense) of the time evolution of the muon
spin polarization~\cite{Uemura_sg,Brewer_usr}.

In the case of a single domain ferromagnet with a unique
crystallographic  muon site with minute dispersion in the internal
magnetization, we would  expect to see coherent muon precession.
However, in an amorphous ferromagnetic material such  as
a$-$Fe$_{1-x}$Mn$_x$, there are several chemically equivalent sites for
the muon, and this wipes away any signature of a muon precession.  With
this in mind, we used the simplest parametrization of our data
consistent with no coherent muon
precession in the ferromagnetic phase  of our samples. Specifically,
we have parametrized our asymmetry data $A(t)$ with a simple dynamical
Kubo-Toyabe relaxation function~\cite{Uemura_sg,Brewer_usr}:
\begin{eqnarray} A(t)    &       =       & A_s(t) \times A_d(t)  \\
A_s(t)  &       =       & \biggl [
                \frac{1}{3} +   \frac{2}{3}  \Bigl \{ 1   - {(
         \Delta t)}^\alpha \Bigr \} \exp\Bigl \{ -\frac{ {(
        \Delta t)}^\alpha}{\alpha} \Bigr \} \biggr ]    \nonumber \\
A_d(t)  &       =       & \biggl [
        \exp \bigl \{ -(\lambda t)^\beta \bigr \}      \biggr] \;\;
        .  \nonumber
        \end{eqnarray} 
\noindent The second term in square brackets for $A_s(t)$ 
describes the $\mu^+$ spin depolarization arising from the static local
magnetic fields statistically 
not parallel to the incoming muon spin at the muon
sites.  Consequently, the long-time limit of the static relaxation function,
$\lim_{t\rightarrow \infty} A_s(t) = 1/3$ for an amorphous ferromagnet,
as occurs  in a spin glass~\cite{Uemura_sg}.
The probability distribution of local fields  is characterized
by a temperature, $T$, dependent RMS field value $\Delta(T)/gamma_{\mu}$,
where $\gamma_{\mu}$ is the muon
gyromagnetic ratio.
$\Delta(T)$ is a measure of the average static electronic moment, $\vec
S_i$, with $\Delta(T) \propto [<\vec S_i>_T^2]^{1/2}$, where $<...>_T$ is a
Boltzman thermal average, and $[...]$ is an average over the spin sites
$i$.  Hence, $\Delta(T)$ is zero in the paramagnetic phase, and
increases as the thermal fluctuations diminish and the spins $\vec S_i$
become increasingly more static.
$A_d(t)$ describes the
fluctuating dynamical local field which leads to a muon spin
depolarization at a rate $\lambda(T)$, where $\lambda(T)$ is an
effective $1/T_1$ spin-lattice relaxation rate.  
$\alpha(T)$ and $\beta(T)$  are temperature
dependent fitting parameters.  
To avoid a non-unique and
an over-parametrization of our data, and
to keep the physical interpretation
of the data as transparent as possible, we analysed our ZF data with
Eq.(1), 
which is qualitatively compatible with our $\mu$SR
spectra for an amorphous ferromagnetic material devoid of muon spin precession
signal. 
Below $T_c$, 
a strong depolarization is observed at early times due to the onset
of static magnetic order.
The ZF spectra were fitted with Eq. 1 using 
$\Delta(T)$, $\lambda(T)$, $\alpha(T)$ and $\beta(T)$ as fitting parameters.

Figures 1a) and 1b)  show the temperature dependence of the  dynamical 
relaxation rate, $\lambda(T)$ of the two reentrant samples, $x=0.26$ 
and 0.30. We observe {\it two} clear peaks at temperatures that correspond 
reasonably well to 
the Curie  temperature, $T_c$, and the onset of strong magnetic 
irreversibility temperature, $T_f$, respectively, as determined by 
magnetization measurements. For the spin glass sample, $x=0.41$, there is a 
single peak in $\lambda$ whose temperature coincides with the spin-glass 
transition temperature $T_f$ found in magnetization measurements (Fig. 1c).  The 
location of the peaks in $\lambda(T)$ for the three samples studied are 
reported in the phase diagram shown in Fig. 1). 

Above $T_c$, the full asymmetry is relaxing dynamically.  Below $T_c$, a static 
intradomain magnetization sets in, and at ``long times'', only 1/3 of the 
asymmetry is relaxing with a dynamical rate $\lambda(T)$. Hence, a 
restricted fit to the long time tail of the asymmetry for
$T<T_c$  reveals the behavior of 
$\lambda(T)$ in the most accurate way.   By analogy with NMR relaxation in 
conventional magnets we attribute most of the observed spin relaxation below 
$T_c$ to muon spin-flip quasi-elastic scattering such as due to two-magnon 
Raman processses, for example. 
For the 
$x=0.26$ and $x=0.30$ 
reentrant samples, we have found that $\lambda(T)$ does not show
{\it any} feature in a 
temperature range around the temperature where previous neutron 
experiments on a$-$Fe$_{1-x}$Mn$_x$ found an anomaly ascribed to a transverse 
spin freezing transition~\cite{Mirebeau}. 
The peak in $\lambda$ seen at $T\approx T_f$, and 
the subsequent rapid decrease in the muon depolarization rate below $T_f$ 
indicate that there is a considerable slowing down of the spin dynamics as the 
system goes through $T_f$, which is then followed by a sizeable reduction in 
the density of low-lying excitations below $T_f$.  Recall that the most naive 
interpretation for the transition at $T_f$ is that of a remnant  of the 
Almeida-Thouless line in an Ising spin glass~\cite{BYFH}.  In this context, 
it is interesting that we see such a dramatic change of spin dynamics 
at a well characterized temperature for which mean-field 
theory predicts no transition~\cite{BYFH}.
This change in the spin dynamics is reminiscent of the anomalous behavior
of the spin wave stiffness constant showing a minimum in
inelastic neutron scattering experiments, and which is unexplained by
theory~\cite{Hennion}. Hence, both 
inelastic neutron scattering and the 
$\mu$SR results show a modification in the nature of the spin excitations
below $T_f$ in a$-$Fe$_{1-x}$Mn$_x$.
For the $x=0.41$ spin glass sample (Fig. 1c),
we have found that the dynamical exponent 
$\beta(T)$ reaches a value very close to 1/3 right 
at $T=T_f$, as observed in recent $\mu$SR experiments on AgMn and 
AuFe dense spin glasses~\cite{Campbell}. 

Figures 2b) and 2c) show the temperature dependence of the static relaxation 
rate, $\Delta(T)$, for the reentrant samples, $x=0.26$ and  $0.30$. 
We expect $\Delta$ to scale like the 
static RMS internal field for $T_{xy}<T<T_c$~\cite{Mirebeau}, with the latter 
itself proportional to the magnetization, $M$.  The solid line in Fig. 2b)
connecting the magnetization data (circles)
shows the expected behavior of $\Delta(T)$ assuming $\Delta\propto M$,
taking the magnetization, $M$, data from Ref.~\cite{Mirebeau}, and fixing an 
overall proportionality factor using the values of $\Delta$ and $M$ at 130K.
The results for $\Delta(T)$ and $M$ depart from each other at
$T\approx T_c \approx 200$K
since $M$
was obtained in a field of 2000Oe, and this has the largest effect close
to $T_c$.
We also find that above 70K, $\Delta(T)$ scale proportionally
to the internal longitudinal field, $B_{\rm ND}$, deduced from
neutron depolarization measurements~\cite{Mirebeau}.
However, $\Delta$ departs from the rescaled $M$ for a 
temperature close to and below
$T=70$K. 
This indicates an increase of the static spin order at $T<T_{xy}\approx 70$K.
A comparison between magnetization data and hyperfine field in
M\"ossbauer data for an $x=0.235$ a-FeMn sample is shown in Fig. (2a).
Overall, the behavior we observe in the static order
parameter  $\Delta$ 
is qualitatively very similar to what is observed in M\"ossbauer 
experiments~\cite{Ryan,Mirebeau} of a-FeMn and other
reentrant spin glasses, where the measured hyperfine field, $B_{\rm Moss}$, 
starts deviating 
{\it slowly} from the bulk magnetization at $T_{xy}$.
However, here in the $\mu$SR experiment, in contrast to most previous
experimental probes used to study reentrant spin glasses, we 
monitor simultaneously
both the static and dynamical behavior  of the magnetic moments. 
\begin{figure}
\begin{center}
  {
  \begin{turn}{0}%
    {\epsfig{file=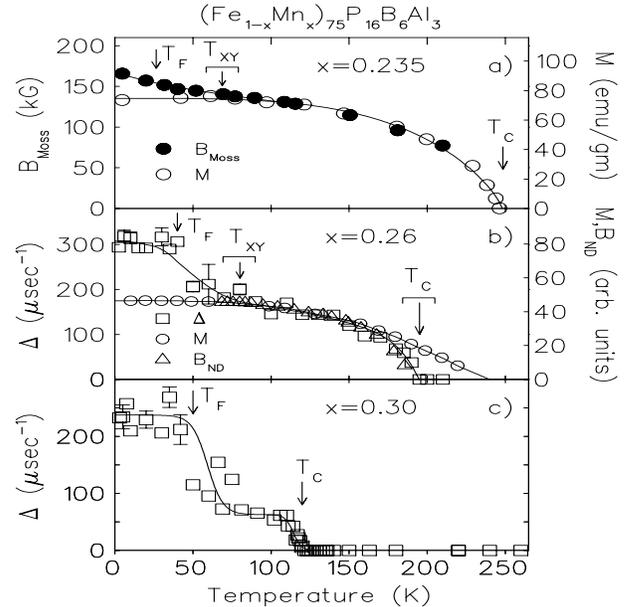,height=8cm,width=8cm} }
   \end{turn}
   }
\caption{Fig. (2a) shows the temperature dependence of the
hyperfine field, $B_{\rm Moss}$,
in a M\"ossbauer
experiment [16] for an $x=0.235$ reentrant a-Fe$_{1-x}$Mn$_x$
(filled circles),
as well as the DC magnetization of that sample, $M$, in arbitrary units
(open circles).
Figs. (2b) and (2c) show the temperature dependence of the static
$\mu$SR relaxation rate, $\Delta(T)$, for the reentrant samples $x=0.26$ and
$x=0.30$, respectively (open squares).
Fig. (2b) also shows the internal field as estimate from neutron depolarization
measurements, $B_{\rm ND}$,  (open triangles) as
as well as the DC magnetization of that sample, $M$, in arbitrary units
(open circles).}
\end{center}
\end{figure}

The above data for $\Delta(T)$ were determined by fitting the 
time dependence of the recovery of $G(t)$ from its minimimum at 
a time $t_{\rm min}=(1+\alpha)^{1/\alpha}/\Delta(T)$ towards the 1/3 value.
In $\mu$SR measurements, there is
a  ``dead time'' intrinsic to the experimental technique, 
which is of the order
of 10ns, and  for which there are no positron counts available.
Setting this dead time to $t_{\rm min}$, we find that for
$\alpha=2$, the largest $\Delta$ we can extract reliably is of
the order of 200$\mu$s$^{-1}$. 
To estimate the value of $\Delta(T=4K)$ independently,
we applied longitudinal fields, $H_{\rm LF}$, and studied the
field dependence of a decoupling ratio
$r \equiv A(t;T=4K)/A(t;T=270K)$ at a time $t \sim 0.05 \mu$s, and fitted $r$ 
using the field-dependence of the Kubo-Toyabe
formula for Gaussian fields~\cite{hayano}.
This is a reasonable procedure since we
have $\Delta=0$ at 270K, $1/\lambda(T=270{\rm K}) \gg (0.05\mu)^{-1}$s and
also $\lambda(T=4 {\rm K}) \gg (0.05\mu)^{-1}$s.
$r$ is given by
\begin{equation}
 r= 1 - \frac{2\Delta^2}{\omega_0^2}+
	\frac{2\Delta^3}{\omega_0^3}
	\exp(-\frac{\omega_0^2}{2\Delta^2})\times
	\int_0^{\omega_0/\Delta}\exp(u^2/2)du
\end{equation}
\noindent 
In the analysis, we assumed that the internal field, 
and hence
$\omega_0=0$ for an applied magnetic field, $H_{\rm LF}$,
$H_{\rm LF} \le  D$, 
and $\omega_0/\gamma_{\mu} = (H_{\rm LF} - D)$ for
$H_{\rm LF} >   D$,
where $D$ is the demagnetization field, and $\gamma_{\mu} =
2\pi \times 13.554 \times 10^7/$Ts is the muon
gyromagnetic ratio.
The sample was bolted in the sample holder to
prevent it from reorienting while in the ferromagnetic phase and
in strong  $H_{\rm LF}$.
The resuls are shown in Fig. 3.
Using $D$ and
$\Delta(T=4{\rm K})$ as fitting parameters, we
found $D=9.23kG$, a reasonable value for a-FeMn for $x=0.26$
for the geometry considered~\cite{Mirebeau},
and $\Delta(T=4{\rm K})/\gamma_{\mu} =0.27$T,
giving $\Delta(T=4{\rm K}) = 230(\mu{\rm s})^{-1}$, in 
reasonable agreement 
with the result obtained by fitting the time dependence of
$A(t)$. In principle, this procedure could be used to extract
$\Delta(T)$ reliably for all $T<T_c$, but would take a prohibitively
long total beamtime.
Hence, we confirm that $\Delta(T\rightarrow 0)$ is above the 
apparent flattening of $\Delta\approx 180(\mu{\rm s})^{-1}$
seen in the range 70K$<$T$<$100K, and
that extra static that is
not parallel to the longitudinal magnetization
develops in the range 50$<T<$100K.
\begin{figure}
\begin{center}
  {
  \begin{turn}{90}%
    {\epsfig{file=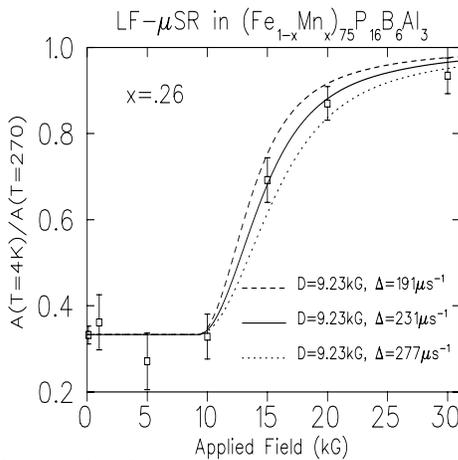,height=6cm,width=6cm} }
   \end{turn}
   }
\caption{Longitudinal field dependence of the asymmetry ratio,
$A(T=4{\rm K})/A(T=270{\rm K})$. The fits are done using the
decoupling Kubo-Toyabe Eq. (2) for the known demagnetization field
$D\approx 9.23$kG for the geometry used in the $\mu$SR experiment, and
for various choices of the static relaxation rate $\Delta$.}
\end{center}
\end{figure}

In conclusion, we have used the muon spin relaxation technique
to study the static and dynamic spin behavior in the amorphous metallic
FeMn to search for {\it three} distinct transitions
in reentrant spin glass materials.
We have found a strong increase of the dynamic as well
as static muon depolarization rate at the paramagnetic to ferromagnetic
transition temperature, $T_c$, and at $T_f$ ($T_f<T_c$) that
corresponds to the onset of strong bulk magnetic irreversabilities. However
we observed no
dynamical (critical slowing down) 
signature of a freezing of the transverse $XY$ spin
components at an intermediate temperature $T_{xy}$ ($T_f < T_{xy}  <
T_c$).  Despite the absence of critical dynamics at
$T_{xy}$,  our  results show
unambiguously that in a reentrant spin glass,
extra static moment which does not contribute to the
magnetization,
develops smoothly with no critical behavior
within the time window of the probe
at a temperature $T_{xy}$
($T_f < T_{xy} < T_c$).
Our results suggest that 
$T_f$ and $T_{xy}$ are due to distinct thermodynamic
``features'' intrinsic to the system, and are not simply arising
from a single transition 
observed at different time and length scales when comparing results
obtained from different experimental techniques.

We acknowledge NSERC (Canada), NEDO (Japan)
and NSF (USA; DMR-95-10454) for financial suppoprt;
S. Barsov, M. Hennion, I. Campbell, and P. Holdsworth  for useful
discussions; and A. Keren and G. Morris for technical
assistance.


\end{document}